\documentclass[12pt]{article}
\usepackage{times}
\usepackage[dvips]{graphicx,color}
\newcommand{\ignore}[1]{}

\newcommand{\mComment}[1]{}
\newcommand{\nComment}[1]{}
\ignore{
\renewcommand{\mComment}[1]{\textcolor{blue}{Manny: #1}}
\renewcommand{\nComment}[1]{\textcolor{magenta}{Michael: #1}}
}

\newcommand{\ket}[1]{|{#1}\rangle}

\newcommand{\tensor}{\otimes}
\newcommand{\trace}{\mbox{tr}}

\newcommand{\crdl}{\|}

\begin{document}

\textsc{\textbf{Quantum information processing, science
    of}}\footnote{Article by E.~H.~Knill and M.~A.~Nielsen accepted
  for Supplement III, Encyclopaedia of Mathematics (publication
  expected Summer 2001).  See also
  ``http://www.wkap.nl/series.htm/ENM''.  E.~H.~Knill is with the Los
  Alamos National Laboratory, MS B265, Los Alamos NM 87545, USA, and
  M.~A.~Nielsen is with the Center for Quantum Computer Technology,
  Department of Physics, University of Queensland 4072, Australia.  }
- The theoretical, experimental and technological areas covering the
use of quantum mechanics for communication and computation. Quantum
information processing includes investigations in quantum information
theory, quantum communication, quantum computation, quantum algorithms
and their complexity, and quantum control.  The science of quantum
information processing is a highly interdisciplinary field. In the
context of mathematics it is stimulating research in pure mathematics
(e.g.  coding theory, $*$-algebras, quantum topology) as well as
requiring and providing many opportunities for applied mathematics.

The science of quantum information processing emerged from the
recognition that usable notions of information need to be physically
implementable. In the 1960s and 1970s researchers such as R.~Landauer,
C.~Bennett, C.~Helstrom and A.~Holevo realized that the laws of
physics give rise to fundamental constraints on the ability to
implement and manipulate information.  Landauer repeatedly stated that
``information is physical'', providing impetus to the idea that it
should be possible to found theories of information on the laws of
physics. This is in contrast to the introspective approach which led
to the basic definitions of computer science and information theory as
formulated by A.~Church, A.~Turing, C.~Shannon and others in the first
half of the 20th century.

Early work in studying the physical foundations of information focused
on the effects of energy limitations and the need for dissipating heat
in computation and communication.  Beginning with S.~Wiesner's work on
applications of quantum mechanics to cryptography in the late 1960s,
it was realized that there may be intrinsic advantages to using
quantum physics in information processing.  Quantum cryptography and
quantum communication in general were soon established as interesting
and non-trivial extensions of classical communication based on
bits. That quantum mechanics may be used to improve the efficiency of
algorithms was first realized when attempts at simulating quantum
mechanical systems resulted in exponentially complex algorithms
compared to the physical resources associated with the system
simulated.  In the 1980s, P.~Benioff and R.~Feynman introduced the
idea of a quantum computer for efficiently implementing quantum
physics simulations.  Models of quantum computers were developed by
D.~Deutsch, leading to the formulation of artificial problems that
could be solved more efficiently by quantum than by classical
computers.  The advantages of quantum computers became widely
recognized when P.~Shor (1994) discovered that they can be used to
efficiently factor large numbers --- a problem believed to be hard for
classical deterministic or probabilistic computation and whose
difficulty underlies the security of widely used public key encryption
methods. Subsequent work established principles of quantum
error-correction to ensure that quantum information processing was
robustly implementable.  See~\cite{Nielsen00a,Gruska99a} for
introductions to quantum information processing and a quantum
mechanics tutorial.

In the context of quantum information theory, information in the sense
of C.~Shannon is referred to as \emph{classical} information. The
fundamental unit of classical information is the \emph{bit}, which can
be understood as an ideal system in one of two states or
configurations, usually denoted by $0$ and $1$. The fundamental units
of quantum information are qubits (short for ``quantum bits''), whose
states are identified with all ``unit superpositions'' of the
classical states. It is common practice to use the bra-ket conventions
for denoting states. In these conventions, the classical
configurations are denoted by $\ket{0}$ and $\ket{1}$, and
superpositions are formal sums $\alpha\ket{0}+\beta\ket{1}$, where
$\alpha$ and $\beta$ are complex numbers satisfying
$|\alpha|^2+|\beta|^2 = 1$.  The states $\ket{0}$ and $\ket{1}$
represent a standard orthonormal basis of a two-dimensional Hilbert
space. Their superpositions are unit vectors in this space. The state
space associated with $n>1$ qubits is formally the tensor product of
the Hilbert spaces of each qubit. This state space can also be
obtained as an extension of the state space of $n$ classical bits by
identifying the classical configurations with a standard orthonormal
basis of a $2^n$ dimensional Hilbert space.

Access to qubit states is based on the postulates of quantum mechanics
with the additional restriction that they are \emph{local} in the
sense that elementary operations apply to one or two qubits at a time.
Most operations can be expressed in terms of standard measurements
of a qubit and two-qubit \emph{quantum gates}.  The standard qubit
measurement has the effect of randomly projecting the state of the
qubit onto one of its classical states; this state
is an output of the measurement (accessible for use in a
classical computer if desired). For example, using the tensor product
representation of the state space of several qubits, a measurement of
the first qubit is associated with the two projection operators
$P^{(1)}_0=P_0\tensor I\tensor\ldots$ and $1-P^{(1)}_0$, where
$P_0\ket{0} = \ket{0}$ and $P_0\ket{1} = 0$.  If $\mathbf{\psi}$ is
the initial state of the qubits, then the measurement outcome is $0$
with probability $p_0=\crdl P_0\mathbf{\psi}\crdl^2$, in which case the new
state is $P_0\mathbf{\psi}/p_0$, and the outcome is $1$ with
probability $1-p_0=\crdl P_1\mathbf{\psi}\crdl^2$ with new state
$P_1\mathbf{\psi}/(1-p_0)$. This is a special case of a
\emph{von~Neumann measurement}.  A general two-qubit quantum gate is
associated with a unitary operator $U$ acting on the state space of
two qubits. Thus $U$ may be represented by a $4\times 4$ unitary
matrix in the standard basis of two qubits.  The quantum gate may be
\emph{applied} to any two chosen qubits. For example, if the state of $n$
qubits is $\mathbf{\psi}$ and the gate is applied to the first two
qubits, then the new state is given by $(U\tensor I\tensor\ldots
)\mathbf{\psi}$. Another important operation of quantum information
processing is preparation of the $\ket{0}$ state of a qubit,
which can be implemented in terms of a measurement and subsequent
applications of a gate depending on the outcome.

Most problems of theoretical quantum information processing can be
cast in terms of the elementary operations above, restrictions on how
they can be used and an accounting of the physical \emph{resources} or
\emph{cost} associated with implementing the operations. Since
classical information processing may be viewed as a special case of
quantum information processing, problems of classical information
theory and computation are generalized and greatly enriched by the
availability of quantum superpositions.  The two main problem areas of
theoretical quantum information processing are quantum computation and
quantum communication.

In studies of quantum computation (cf. \textbf{quantum computation})
one investigates how the availability of qubits can be used to improve
the efficiency of algorithmic problem solving. Resources counted
include the number of quantum gates applied and the number of qubits
accessed. This can be done by defining and investigating various types
of quantum automata, most prominently quantum Turing machines, and
studying their behavior using approaches borrowed from the classical
theory of automata and languages. It is convenient to combine
classical and quantum automata, for example by allowing a classical
computer access to qubits as defined above, and then investigating the
complexity of algorithms by counting both classical and quantum
resources, thus obtaining trade-offs between the two.

Most of the complexity classes for classical computation have
analogues for quantum computation, and an important research area is
concerned with establishing relationships between these complexity
classes.
Corresponding to the classical class $\mathbf{P}$ of polynomially
decidable languages is the class of languages decidable in bounded
error quantum polynomial time, $\mathbf{BQP}$. While it is believed
that $\mathbf{P}$ is properly contained in $\mathbf{BQP}$, whether
this is so is at present an open problem.  $\mathbf{BQP}$ is known to
be contained in the class $\mathbf{P}^{\raisebox{.5pt}{\#}\mathbf{P}}$
(languages decidable in classical polynomial time given access to an
oracle for computing the permanent of $0$-$1$ matrices), but the
relationship of $\mathbf{BQP}$ to the important class of
nondeterministic polynomial time languages $\mathbf{NP}$ is not known.

In quantum communication one considers the situation where two or more
entities with access to local qubits can make use of both classical
and quantum (communication) channels for exchanging information. The
basic operations now include the ability to send classical bits and
the ability to send quantum bits.  There are two main areas of
investigation in quantum communication.  The first aims at determining
the advantages of quantum communication for solving classically posed
communication problems with applications to
cryptography and to distributed computation. The second is concerned
with establishing relationships between different types of
communication resources, particularly with respect to noisy quantum
channels, thus generalizing classical communication theory. 

Early investigations of quantum channels focused on using them for
transmitting classical information by encoding a source of information
(cf. \textbf{information, source of}) with uses of a quantum channel
(cf. \textbf{quantum communication channel}).  The central result of
these investigations is A.~Holevo's bound (1973) on the amount of
classical information that can be conveyed through a quantum
channel. Asymptotic achievability of the bound (using block coding of
the information source) was shown in the closing years of the
twentieth century.
With some technical caveats, the bound and its
achievability form a quantum information-theoretic analogue of
Shannon's capacity theorem for classical communication channels.

Quantum cryptography, distributed quantum computation and quantum
memory require transmitting (or storing) quantum states.  As a result
it is of great interest to understand how one can communicate quantum
information through quantum channels.  In this case, the source of
information is replaced by a source of quantum states, which are to be
transmitted through the channel with high \emph{fidelity}.  As in the
classical case, the state is encoded before transmission and decoded
afterwards.  There are many measures of fidelity which may be used to
evaluate the quality of the transmission protocol. They are chosen so
that a good fidelity value implies that with high probability, quantum
information processing tasks behave the same using the original or the
transmitted states. A commonly used fidelity measure is the
Bures-Uhlmann fidelity, which is an extension of the Hilbert space
norm to probability distributions of states (represented by
\emph{density operators}). In most cases, asymptotic properties of
quantum channels do not depend on the details of the fidelity measure
adopted.

To improve the reliability of transmission over a noisy quantum
channel, one uses \emph{quantum error-correcting codes} to encode a
state generated by the quantum information source with multiple uses
of the channel.  The theory of quantum codes can be viewed as an
extension of classical coding theory. Concepts such as minimum
distance and its relationship to error-correction generalize to
quantum codes.  Many results from the classical theory, including some
linear programming upper bounds and the Gilbert-Varshamov lower bounds
on the achievable rates of classical codes have their analogues for quantum
codes. In the classical theory, linear codes are particularly useful
and play a special role. In the quantum theory, this role is played by
the \emph{stabilizer} or \emph{additive} quantum codes, which are in
one-to-one correspondence with self-dual (with respect to a specific
symplectic inner product) classical $\mathrm{GF}_2$-linear codes over
$\mathrm{GF}_4$ (cf. \textbf{finite fields}).

The capacity of a quantum channel with respect to encoding with
quantum codes is not as well understood as the capacity for
transmission of classical information.  The exact capacity is known
only for a few special classes of quantum channels. Although there are
information theoretic upper bounds, they depend on the number of
channel instances, and whether or not they can be achieved is an open
problem.  A further complication is that the capacity of quantum
channels depends on whether one-way or two-way classical communication
may be used to restore the transmitted quantum
information~\cite{bennett:qc1996a}.

The above examples illustrate the fact that there are many different
types of information utilized in quantum information theory, making it
a richer subject than classical information theory.  Another physical
resource whose properties appear to be best described by
information-theoretic means is {\em quantum entanglement}.  A quantum
state of more than one quantum system (e.g. two qubits) is said to be
entangled if the state can not be factorized as a product of states of
the individual quantum systems.  Entanglement is believed to play a
crucial role in quantum information processing, as demonstrated by its
enabling role in effects such as quantum key distribution, superdense
coding, quantum teleportation, and quantum error-correction.
Beginning in 1995 an enormous amount of effort has been devoted to
understanding the principles governing the behavior of entanglement.
This has resulted in the discovery of connections between quantum
entanglement and classical information theory, the theory of positive
maps~\cite{Horodecki96f} and majorization~\cite{Nielsen99a}.

The investigation of quantum channel capacity, entanglement, and many
other areas of quantum information processing involves various quantum
generalizations of the notion of entropy, most notably the von~Neumann
entropy. The von~Neumann entropy is defined as
$H(\rho)=\trace\rho\log_2(\rho)$ for density operators $\rho$ ($\rho$
is positive Hermitian and of trace $1$). It has many (but not all) of
the properties of the classical information function $H(\cdot)$
(cf. \textbf{information, amount of}). Understanding these properties
has been crucial to the development of quantum information processing
(see~\cite{Nielsen00a,Wehrl78a,Ohya93a} for reviews).  Probably the
most powerful known result about the von~Neumann entropy is the strong
subadditivity inequality.
Many of the bounds on quantum communication follow as easy corollaries
of strong subadditivity. Whether still more powerful entropic
inequalities exist is not known.

An important property of both classical and quantum information is
that although it is intended to be physically realizable, it is
abstractly defined and therefore independent of the details of a
physical realization. It is generally believed that qubits encapsulate
everything that is finitely realizable using accessible physics. This
belief implies that any information processing implemented by
available physical systems using resources appropriate for those
systems can be implemented as efficiently (with at most polynomial
overhead) using qubits. It is noteworthy that there is presently no
proof that information processing based on quantum field theory (cf.
\textbf{quantum field theory}) is not more efficient than information
processing with qubits.  Furthermore, the as-yet unresolved problem of
combining quantum mechanics with general relativity in a theory of
quantum gravity prevents a fully satisfactory analysis of the
information processing power afforded by fundamental physical laws.

Much effort in the science of quantum information processing is being
expended on developing and testing the technology required for
implementing it. An important task in this direction is to establish
that quantum information processing can be implemented robustly in the
presence of noise. At first it was believed that this was not
possible. Arguments against the  robustness of quantum information
were based on the apparent relationship to analogue computation (due
to the continuity of the amplitudes in the superpositions of
configurations) and the fact that it seemed difficult to observe
quantum superpositions in nature (due to the rapid loss of phase
relationships called \emph{decoherence}). However, the work on quantum
error-correcting codes rapidly led to the realization that provided
the physical noise behaves locally and is not too large, it is at
least in principle possible to process quantum information fault
tolerantly. Research in how to process quantum information reliably
continues; the main problems is improving the
estimates on the maximum amount of tolerable noise for general models
of quantum noise and for the types of noise expected in specific
physical systems. Other issues include the need to take into
consideration restrictions imposed by possible architectures and
interconnection networks.

There are many physical systems that can potentially be used for
quantum information processing~\cite{braunstein:qc2000b}. An active
area of investigation involves determining the general mathematical
features of quantum mechanics required for implementing quantum
information.  More closely tied to existing experimental techniques
are studies of specific physical systems.  In the context of
communication, optical systems are likely to play an important role,
while for computation there are proposals for using electrons or
nuclei in solid state, ions or atoms in electromagnetic traps,
excitations of superconductive devices etc.  In all of these,
important theoretical issues arise. These issues include how to
optimally use the available means for controlling the quantum systems
(\emph{quantum control}), how to best realize quantum information
(possibly indirectly), what architectures can be implemented, how to
translate abstract sequences of quantum gates to physical control
actions, how to interface the system with optics for communication,
refining the theoretical models for how the system is affected by
noise and thermodynamic effects, and  how to reduce the effects of noise.


\begin{center}\mbox{}\hspace*{\fill}
\begin{tabular}{l}
E. H. Knill\\
M. A. Nielsen
\end{tabular}
\end{center}

\noindent\textsf{AMS 2000 Subject Classification: 81P68, 68Q05}

\end{document}